\documentstyle[epsf]{mn}
\newif\ifAMStwofonts
\AMStwofontstrue

\ifoldfss
  \ifCUPmtlplainloaded \else
    \NewTextAlphabet{textbfit} {cmbxti10} {}
    \NewTextAlphabet{textbfss} {cmssbx10} {}
    \NewMathAlphabet{mathbfit} {cmbxti10} {} 
    \NewMathAlphabet{mathbfss} {cmssbx10} {} 
  \fi
  \ifAMStwofonts
    \ifCUPmtlplainloaded \else
      \NewSymbolFont{upmath} {eurm10}
      \NewSymbolFont{AMSa} {msam10}
      \NewMathSymbol{\upi}     {0}{upmath}{19}
      \NewMathSymbol{\umu}     {0}{upmath}{16}
      \NewMathSymbol{\upartial}{0}{upmath}{40}
      \NewMathSymbol{\leqslant}{3}{AMSa}{36}
      \NewMathSymbol{\geqslant}{3}{AMSa}{3E}

       \let\le=\leqslant
       \let\ge=\geqslant
    \fi
  \fi
\fi 

\ifnfssone
  \newmathalphabet{\mathit}
  \addtoversion{normal}{\mathit}{cmr}{m}{it}
  \addtoversion{bold}{\mathit}{cmr}{bx}{it}
  \newmathalphabet{\mathbfit} 
  \addtoversion{normal}{\mathbfit}{cmr}{bx}{it}
  \addtoversion{bold}{\mathbfit}{cmr}{bx}{it}
  \newmathalphabet{\mathbfss} 
  \addtoversion{normal}{\mathbfss}{cmss}{bx}{n}
  \addtoversion{bold}{\mathbfss}{cmss}{bx}{n}
  \ifAMStwofonts
    \ifCUPmtlplainloaded \else
      \UseAMStwoboldmath
      \makeatletter
      \new@mathgroup\upmath@group
      \define@mathgroup\mv@normal\upmath@group{eur}{m}{n}
      \define@mathgroup\mv@bold\upmath@group{eur}{b}{n}
      \edef\UPM{\hexnumber\upmath@group}
      \new@mathgroup\amsa@group
      \define@mathgroup\mv@normal\amsa@group{msa}{m}{n}
      \define@mathgroup\mv@bold\amsa@group{msa}{m}{n}
      \edef\AMSa{\hexnumber\amsa@group}
      \makeatother
      \mathchardef\upi="0\UPM19
      \mathchardef\umu="0\UPM16
      \mathchardef\upartial="0\UPM40
      \mathchardef\leqslant="3\AMSa36
      \mathchardef\geqslant="3\AMSa3E

       \let\le=\leqslant
       \let\ge=\geqslant
    \fi
  \fi
\fi 

\ifnfsstwo
  \DeclareMathAlphabet{\mathbfit}{OT1}{cmr}{bx}{it}
  \SetMathAlphabet\mathbfit{bold}{OT1}{cmr}{bx}{it}
  \DeclareMathAlphabet{\mathbfss}{OT1}{cmss}{bx}{n}
  \SetMathAlphabet\mathbfss{bold}{OT1}{cmss}{bx}{n}
  \ifAMStwofonts
    \ifCUPmtlplainloaded \else
      \DeclareSymbolFont{UPM}{U}{eur}{m}{n}
      \SetSymbolFont{UPM}{bold}{U}{eur}{b}{n}
      \DeclareSymbolFont{AMSa}{U}{msa}{m}{n}
      \DeclareMathSymbol{\upi}{0}{UPM}{"19}
      \DeclareMathSymbol{\umu}{0}{UPM}{"16}
      \DeclareMathSymbol{\upartial}{0}{UPM}{"40}
      \DeclareMathSymbol{\leqslant}{3}{AMSa}{"36}
      \DeclareMathSymbol{\geqslant}{3}{AMSa}{"3E}

       \let\le=\leqslant
       \let\ge=\geqslant
    \fi
  \fi
\fi 

\ifCUPmtlplainloaded \else
  \ifAMStwofonts \else 
    \def\upi{\pi}
    \def\umu{\mu}
    \def\upartial{\partial}
  \fi
\fi

\title{Spherical Mexican Hat wavelet: an application to detect
non-Gaussianity in the COBE-DMR maps}
\author[] {L. Cay\'on$^{1}$, J.L. Sanz$^{1}$, 
E. Mart\'\i nez-Gonz\'alez$^{1}$, A. J. Banday$^{2}$,
F. Arg\"ueso$^{3}$,   \\
\newauthor
J.E. Gallegos$^{1}$ 
K. M. G\'{o}rski$^{4,5}$
and G. Hinshaw$^{6}$.\\ 
1. Instituto de F\'\i sica de Cantabria, Fac. Ciencias, Av. los
	Castros s/n, 39005 Santander, Spain\\
2. Max-Planck Institut fuer Astrophysik (MPA), Karl-Schwarzschild Str.1, D-85740, Garching, Germany\\
3. Dpto. de Matem\'aticas, Universidad de Oviedo, c/ Calvo Sotelo s/n, 33007 Oviedo, Spain\\
4. European Southern Observatory (ESO), Karl-Schwarzschild Str.2,
D-85740, Garching, Germany\\
5. Warsaw University Observatory, Poland\\
6. NASA/GSFC, Greenbelt, MD 20771, USA.\\}
\date{\today}

\pagerange{\pageref{firstpage}--\pageref{lastpage}}
\pubyear{2001}

\begin{document}

\maketitle

\label{firstpage}

\begin{abstract}

The spherical Mexican Hat wavelet is introduced
in this paper, with the aim of testing the Gaussianity of the Cosmic Microwave
Background temperature fluctuations. Using the information given by the wavelet
coefficients at several scales, we have performed several statistical
tests on the COBE-DMR maps to search for evidence of non-Gaussianity. Skewness, kurtosis,
scale-scale correlations (for two and three scales) 
and Kolmogorov-Smirnov tests indicate that 
the COBE-DMR data are consistent with a Gaussian distribution. 
We have extended the analysis
to compare temperature values provided 
by COBE-DMR data with distributions (obtained 
from Gaussian simulations) at each pixel and at each scale. The
number 
of pixels with temperature values outside the $95\%$ and the $99\%$ is 
consistent with that obtained from Gaussian simulations, at
all scales. 
Moreover, the extrema values for COBE-DMR data (maximum and minimum
temperatures in the map) are also consistent with those 
obtained from Gaussian simulations.

\end{abstract}


\section{Introduction}

Cosmic Microwave Background (CMB) temperature fluctuations are predicted
to be Gaussian by standard Inflationary theories. 
Topological Defect models and non-standard Inflation can
produce non-Gaussian CMB temperature fluctuations.
The most recent results of CMB observations by Boomerang, DASI and
MAXIMA-I (Netterfield et al. 2001, Pryke et al.
2001, Stompor et al. 2001) seem to indicate that any possible non-Gaussianity present in the 
data will more likely be produced by non-standard Inflationary models.  
Determination of
the possible degree of non-Gaussianity of the temperature fluctuations 
in CMB maps is therefore a fundamental question to be answered. 
Several tests have been proposed on real space ranging from calculating
skewness and kurtosis (Luo \& Schramm 1993a), 3-point correlation functions
(Luo \& Schramm 1993b, Kogut et al 1996), Minkowski
functionals (Gott et al. 1990, Torres et al. 1995, Kogut et al. 1996, 
Schmalzing \& G\'{o}rski 1997, Novikov et al. 2000) and topological
quantities related to characteristics of
excursion sets above a given threshold (Bond \& Efstathiou 1987, Vittorio \& Juzkiewicz 1987, Mart\'\i nez-Gonz\'alez \& Sanz 1989, Barreiro et al. 1997,1998,1999), to multifractal analysis, partition
function (Pompilio et al. 1995, Diego et al. 1999) and surface rougness based analysis (Mollerach et al. 1999). Using the expansion of temperature fluctuations in spherical harmonics, one can test the non-Gaussianity of the CMB by calculating
the bispectrum (Luo 1994, Heavens 1998, Ferreira et al. 1998, Magueijo 2000). Moreover, several recent papers have appeared proposing
analysis in wavelet space (Pando, Valls-Gabaud \& Fang 1998, Hobson, Jones \& Lasenby 1999, Barreiro et al. 2000, Aghanim, Forni \& Bouchet 2000). 
 
Many of the works cited above aimed to detect non-Gaussianity in the 
4-year COBE-DMR data,
which currently remains the only publicly available full-sky CMB data set.
Ferreira, Magueijo \& G\'{o}rski (1998), using a reduced bispectrum that
did not consider correlations between different scales, 
detected a non-Gaussian signal which was most likely produced
by a systematic artifact in the COBE-DMR data
(Banday, Zaroubi \& G\'{o}rski 2000). No detection of non-Gaussianity was
obtained after
that artifact was removed. 
However, non-Gaussianity has been detected at the $97\%-99.8\%$
level by using an extended bispectrum that takes into account 
correlations between different scales by Magueijo (2000). 

Non-Gaussianity has not been revealed by any of the other tests applied 
by different authors,
including those carried out in wavelet space
(Mukherjee, Hobson \& Lasenby 2000, 
Aghanim, Forni \& Bouchet 2000). 
Although the COBE-DMR data cover the 
whole celestial sphere, most of these works studied the wavelet 
coefficient distributions corresponding to pixels covering Faces 0 and 5 
in the QuadCube pixelisation (White \& Stemwedel 1992) . Therefore the 
wavelet coefficients are calculated on the plane. The first attempt to 
detect non-Gaussianity in the COBE-DMR maps analysing all the available
pixels and therefore using a wavelet projected on
the surface of the sphere was presented by Barreiro et al. (2000). The wavelet
used in that work was the spherical Haar wavelet (as introduced to 
CMB analysis by Tenorio et al. 1999). Skewness, kurtosis and two-scale
correlation functions 
were computed in wavelet space corresponding to different scales 
for the 4-year COBE-DMR data. The tests could not confirm any strong
evidence of non-Gaussianity.

Our aim in this paper is to attempt to 
detect non-Gaussianity in the 4-year COBE-DMR data 
by applying different statistical tests to 
the wavelet coefficients obtained by convolving the COBE-DMR map
with a stereographic projection of the 
Mexican Hat wavelet on the sphere. 
This is the first time that the
Mexican Hat wavelet has been used in this projection. The definition and 
basic properties of this wavelet, in particular in relation to its 
application on the sphere, are presented in Section 2 
(an extended discussion will be given in 
Mart\'\i nez-Gonz\'alez et al. 2001). 

It should be noted that,
while the Haar wavelet privileges
some directions (with diagonal, vertical and horizontal components),
the Mexican Hat wavelet is isotropic. We will compare the
ability of both wavelets to detect non-Gaussian features in the 
COBE-DMR data. The method and tests applied in this paper are 
presented in Section 3. Section 4 is dedicated to present and 
discuss the results. Conclusions are also included in that section.

\setcounter{figure}{0}
\begin{figure*}
 \epsffile{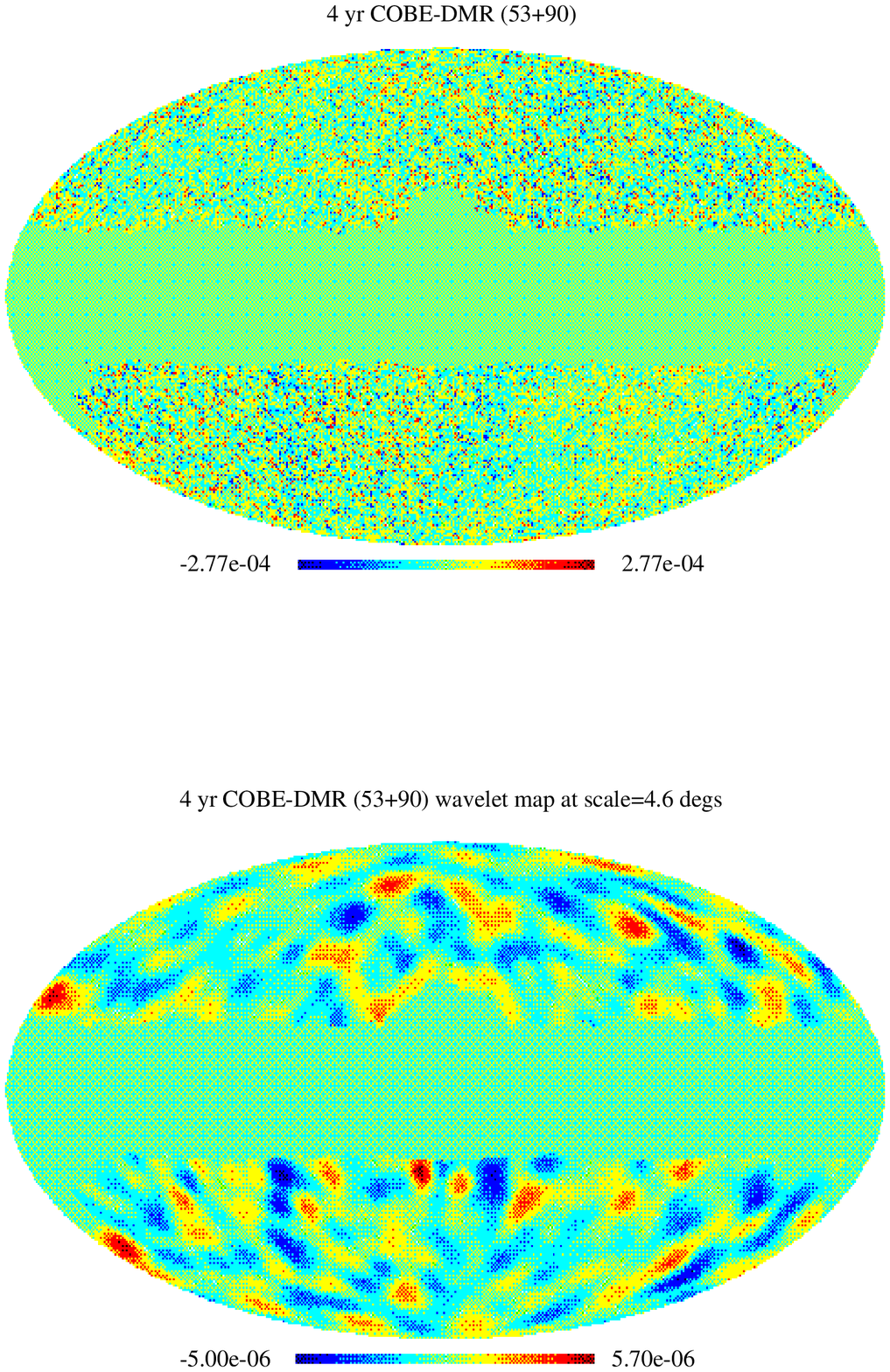}
 \caption{Analysed co-added COBE-DMR map (top) 
and the corresponding wavelet coefficient map obtained at 
$4.6$ degrees (bottom).
}
 \label{f1}
\end{figure*}


\setcounter{figure}{1}
\begin{figure*}
 \epsffile{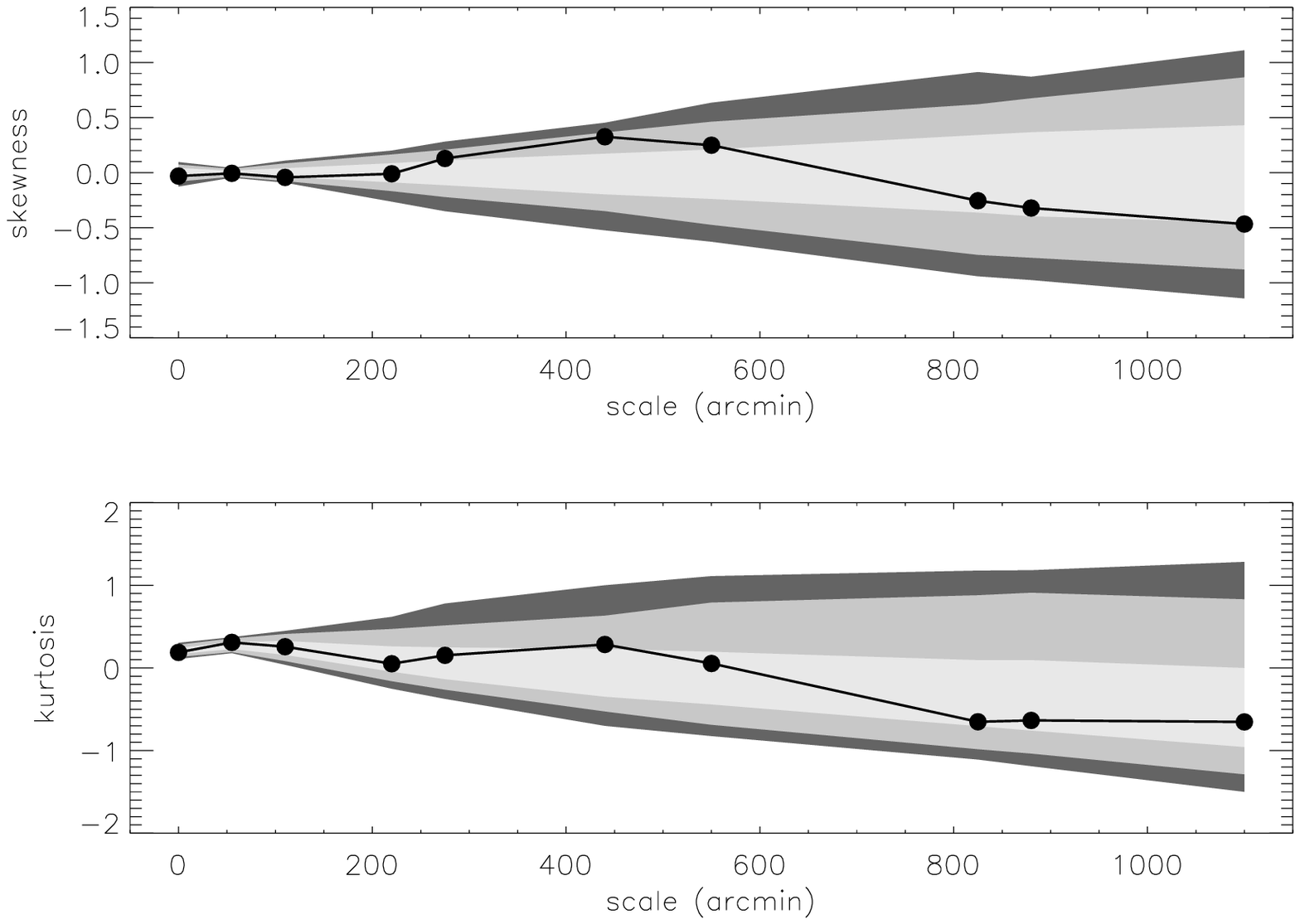}
 \caption{
Skewness values obtained from the analysed co-added COBE-DMR map at
several scales are represented by dots in the top figure. The bands 
represent the $68\%$, $95\%$ and $99\%$ confidence levels. The 
corresponding kurtosis results are presented 
in the 
bottom figure. Values at zero correspond to real space.
}
 \label{f1}
\end{figure*}


\section{The Spherical Mexican Hat Wavelet}

In many cases of astrophysical interest, data are given on the
sphere. In this paper,
we are interested in the analysis of the COBE-DMR data 
in the HEALPix pixelisation\footnote{http://www.eso.org/kgorski/healpix/}
(G\'orski, Hivon \& Wandelt 1999),
but current and future experiments measuring CMB anisotropy can also be 
investigated using these methods. A conventional analysis technique is based on
spherical harmonic decomposition of the temperature anisotropy 
for which local information is
lost. On the other hand, wavelets -defined on the line/plane- have been extensively used 
(spectral and image analysis) in astrophysical applications during the last years. In
particular, the continous isotropic wavelet transform of a 2D signal $f(\vec{x})$ is 
defined as

\begin{equation}
w(R, \vec{b}) = \int d\vec{x}\,f(\vec{x})\Psi (R, \vec{b}; \vec{x}),
\end{equation}
\noindent being
\begin{equation}
\Psi (R, \vec{b}; \vec{x}) = \frac{1}{R}\psi (\frac{|\vec{x}- \vec{b}|}{R}),
\end{equation}
\noindent and where $w(R,\vec{b})$ is the wavelet coefficient associated to the scale $R$ at
the point with coordinates $\vec{b}$. The limits in the double integral are $-\infty$ and
$\infty$ for the two variables. $\psi$ is the wavelet ¨mother¨ function that satisfies the
constraints $\int d\vec{x}\,\psi = 0, \int d\vec{x}\,{\psi}^2 = 1$, and the admissibility condition
$C_{\psi}\equiv (2\pi)^2\int dk\,k^{-1}\,{\psi}^2(k) < \infty $, where $\psi (k), k\equiv |\vec{k}|$, 
is the Fourier transform of $\psi (x)$. A reconstruction of the image can be achieved with 
the inversion formula

\begin{equation}
f(\vec{x}) = \frac{1}{C_{\psi}}\int dR\,d\vec{b}\frac{1}{R^4}w(R, \vec{b})
\psi (\frac{|\vec{x}- \vec{b}|}{R}).
\end{equation}

The analysis of 2D images can be done with the so-called Mexican Hat wavelet 
(MEXHAT). It is given by

\begin{equation}
\psi (x) = \frac{1}{{(2\pi)}^{1/2}}(2 - x^2)e^{-x^2/2},\ \ \  x\equiv |\vec{x}|,
\end{equation}

\noindent that is proportional to the laplacian of a Gaussian. It has been sucessfully 
applied to extract point sources from CMB maps (Cay\'on et al. 2000, Vielva et al. 2000).

The problem arises when one tries to extend the continous isotropic wavelet to the sphere.
There had been some proposals to do such an extension (e. g. analyzing functions defined 
through spherical harmonics, Freeden 1997; second generation wavelets, Sweldens 1996). 
However, one would like to incorporate the following properties:
i) the transform is defined by a wavelet, i. e. $\int d\Omega {\Psi }_{S_2} = 0$, ii) the wavelet 
transform must involve translations and dilations on the sphere and iii) for
small scales the transform must reduce to the usual continous isotropic wavelet on the plane.
Recently, Antoine \& Vandergheynst (1998) have presented a group-theoretical derivation of the
continous wavelet transform on the 2-sphere $S_2$ satisfyng the three previous conditions. They 
identify the appopriate transformations as: i) translations are given by      
elements of the 3D rotation group $SO(3)$; ii) dilations around any point are the inverse
stereographic projection of standard dilations on the tangent plane to the point. Using polar
coordinates $(\theta , \phi )$ around any fixed -but arbitrary- point on $S_2$,
such a projection is defined by

\begin{equation}
x = 2\tan \frac{\theta }{2},\ \ \  x\equiv |\vec{x}|,
\end{equation}

\noindent where $(x, \phi )$ are polar coordinates on the tangent plane. Moreover, the
analyzing wavelet, with respect to any fixed but arbitrary point, is given by 
(Mart\'\i nez-Gonz\'alez et al. 2001)

\begin{equation}
{\Psi }_{S_2}(\theta ; R) = N\frac{4}{{(1 + \cos \theta )}^2}
\psi (\frac{x}{R} = \frac{2}{R}\tan \frac{\theta }{2}),
\end{equation}
\noindent where $N$ is a normalization constant given by
\begin{equation} 
N= {1\over R}{(1 + \frac{R^2}{2} + \frac{R^4}{4})}^{-1/2},
\end{equation}
\noindent being $\psi (x/R)$ the mother wavelet 
associated to the continous transform on the plane. $R$ defines the 
size of the Mexican Hat. Then, the generalization to $S_2$ of the 
MEXHAT on the plane is clearly given by eq. (6) with $\psi (x/R)$ given by eq. (4).
Therefore, moving any point to the North Pole and taking polar coordinates, the wavelet
coefficients associated to the function $f(\theta , \phi )$ are given 
at that point by

\begin{equation}
w(R) = \int d\Omega f(\theta , \phi ){\Psi }_{S_2}(\theta ; R).
\end{equation}

\section{Method} 

The methodology used in this work is designed to analyse the 4-year
COBE-DMR data in the HEALPix pixelisation. Maps at 53 and 90 GHz are
added using inverse-noise-variance weights.
The resolution of these maps is $N_{side}=64$ corresponding
to a pixel size of $\approx 55'$ and resulting in 49152 pixels.
Only those pixels surviving
the extended Galactic cut of Banday et al. (1997) are considered
(in order to excise those pixels with significant
Galactic contamination near the plane of the Galaxy). 
Moreover, we subtract the best-fit monopole and dipole 
(as computed on the cut-sky) from the remaining pixels\footnote{Note 
that this differs slightly from the treatment by Barreiro et al. It has been
checked that in the formalism used in that paper, removing monopole
and dipole does not affect the results}. 
We refer to the resulting map as the co-added COBE-DMR map. 

The spherical Mexican Hat wavelet of several sizes 
is convolved with the co-added COBE-DMR map. The size of the Mexican Hat ($R$)
fixes the characteristic scale  of the wavelet coefficients (obtained after
the convolution). 
Wavelet coefficient maps are 
obtained for several Mexican Hat sizes and statistical 
tests presented below are applied to these maps (see Figure 1 for 
an example of a wavelet coefficients map). Confidence 
intervals are estimated from simulations. We simulate CMB temperature 
fluctuations assuming a CDM power spectrum (computed using the CMBFAST
software) with $\Omega_b=0.05$, $\Omega_c=0.3$, 
$\Omega_v=0.65$, $\Omega_{\nu}=0$, $H_o=65$ km/s/Mpc and $n=1$. The power
spectrum is normalized to $Q_{rms-PS}=18 \mu K$. Simulated maps are smoothed
with a Gaussian beam of $FWHM=7.032^{\circ}$ (to approximate the beam
response of the DMR instrument). Noise maps, generated with the same 
observation pattern and rms noise levels as the COBE-DMR data, are added to the 
simulated signal. The Galactic mask is also applied to the simulations,
and the monopole and dipole components are estimated and subtracted
outside the excluded region for each simulated map.
Since the Mexican Hat wavelet extends to the whole sphere, convolution with 
the sky at a fixed pixel will involve information from pixels over all
angular separations. However, in this analysis the wavelet coefficients 
must be computed excluding those pixels within the Galactic mask.
In principle, this effect could bias the distribution we want to test.
In practise, there is no
appreciable change in results introduced by the partial sky coverage.

We have applied several tests in wavelet space that can be divided into
two categories. The first ones are computed globally from
wavelet coefficient maps at scales ranging from $55'$ to $1760'$. We have also
computed these tests on the real map and results are indicated 
by scale $0'$ in Figure 2 and Table 1.
In addition we have also
considered tests based on the derived distribution function at each
pixel, at scales ranging from $108'$ to $1728'$. In this way we take into 
account the different amplitude of the noise at each pixel.
The reason for considering
scales different from the ones used in the previous case, 
is that the maximum resolution
of the analysed COBE-DMR map and of the simulations is $N_{size}=32$ (corresponding to 12228 pixels on the sphere with size $1^{\circ}.8$). 
Information on temperature distributions at each pixel has to be kept 
at each scale. 
We needed to degrade the resolution to be able to handle all this information.

In the first category of tests we calculated different statistics commonly
used to detect non-Gaussianity. Skewness and kurtosis 
will test the third ($\kappa _3$) and fourth ($\kappa _4$) order cumulants :
\begin{equation}
S=\kappa _3 / \kappa _2^{3/2}\ \ \
K=\kappa_4 /\kappa_2^{2} -3,
\end{equation}
\noindent Correlations between two consecutive scales $k,j$ are tested using the 
conventional 2-scale correlation given by:
\begin{equation}
C_{2,1}(k,j)={{\sum_{i=1}^{N_p}\Delta T/T(k,i)\Delta T/T(j,i)}\over {N_pV(k)V(j)}}
\end{equation}
\noindent where
\begin{equation}
V(k)=({{\sum_{i=1}^{N_p}(\Delta T/T(k,i))^2}\over {N_p}})^{1/2}
\end{equation}
\noindent and $N_p$ will correspond to the number of pixels considered at
the largest scale (the ones outside the Galactic mask located at
a distance from the border larger than twice the Mexican Hat size corresponding
to that scale). As well as a correlation analogous to the one used by Barreiro et al.
2000, given by:
\begin{equation}
C_{2,2}(k,j)={{\sum_{i=1}^{N_p}(\Delta T/T(k,i))^2(\Delta T/T(j,i))^2}\over 
{V(k)V(j)}}
\end{equation}
\noindent We have also considered correlations between three consecutive 
scales $k,j,l$ given by:
\begin{equation}
C_{3}(k,j,l)={{\sum_{i=1}^{N_p}\Delta T/T(k,i)\Delta T/T(j,i)\Delta T/T(l,i)}\over {N_pV(k)V(j)V(l)}}
\end{equation}
\noindent In addition we compare the temperature fluctuation distributions 
from the data to those obtained from simulations, at several scales. The
Kolmogorov-Smirnov test is used in this comparison. We have calculated the
distance parameter for the COBE-DMR data as well as a distance parameter
distribution from simulations.

In the second category of tests we have calculated wavelet coefficient 
distributions at each pixel at each scale. We compute the probability of 
extrema taking into account the dispersion at each pixel.
The second of these tests consists in the following. At a fixed scale, 
for the COBE-DMR data we select those pixels with values outside the $99\%$ and
$95\%$. The number of those selected pixels is compared with the expected
number from simulations.  


\setcounter{figure}{2}
\begin{figure*}
 \epsffile{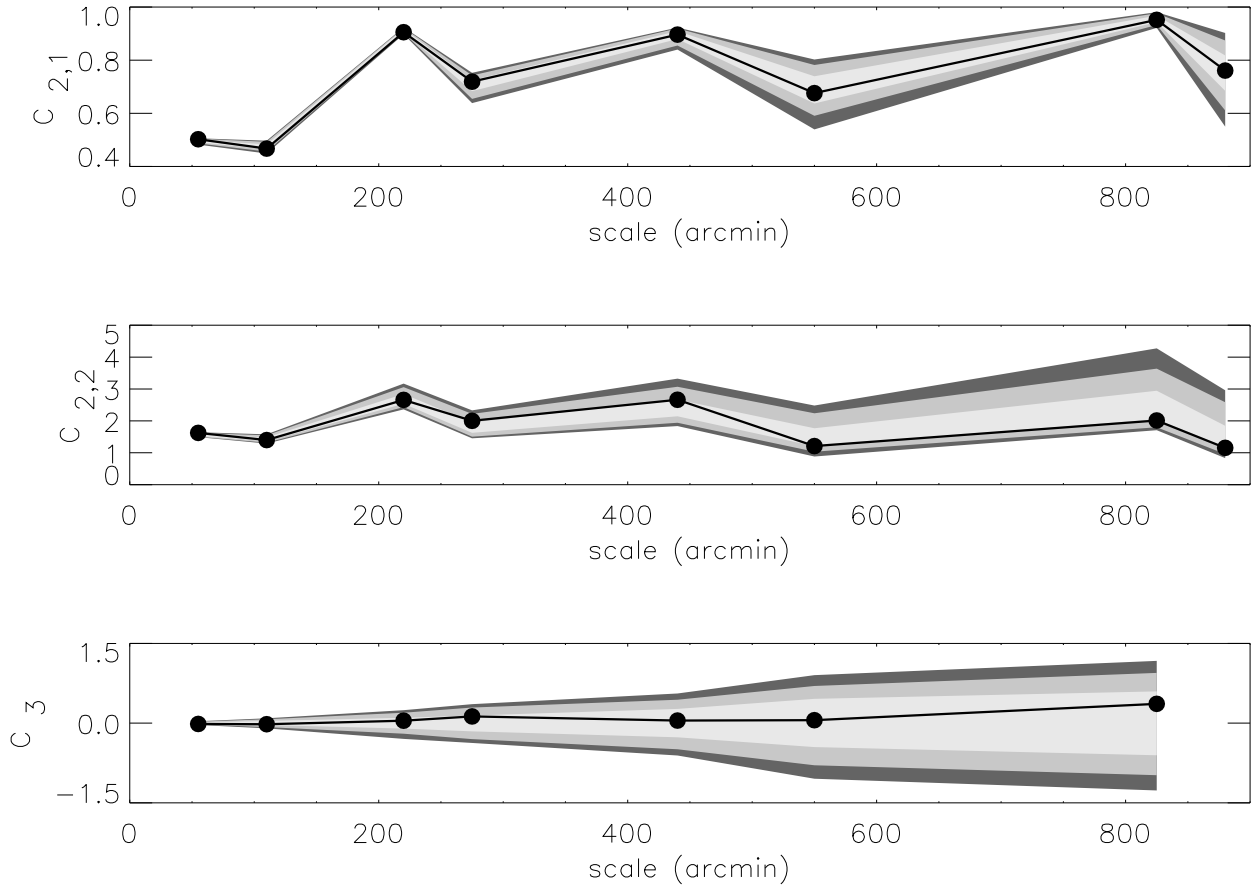}
 \caption{Values corresponding to the 2-scale correlation 
$C_{2,1}$ are presented in the top figure. At each scale, the correlation
has been calculated with the scale directly above (i.e. the value at $55'$
corresponds to the correlation between wavelet coefficients
at $55'$ and $110'$). The scales 
are the same ones appearing in Table 1, from $55'$ to $1100'$. 
Bands correspond to confidence levels at $68\%$, $95\%$ and $99\%$. The middle figure presents the results for the 2-scale 
correlation defined as in Barreiro et al. 2000 $C_{2,2}$. The figure
in the bottom corresponds to the correlation between three consecutive
scales. At each scale the value presented is the correlation between
that scale and the other two direclty above that one (i.e. at $55'$ 
the correlation is calculated between wavelet coefficients at that
scale and $110'$ and $220'$).}
 \label{f1}
\end{figure*}





\begin{table*}
  \begin{minipage}{170mm}
  \begin{center}
  \caption[]{Kolmogorov-Smirnov distance dKS}
  \label{tab1}
  \begin{tabular}{c|c|c}\hline
  scale(') & dKS(COBE)& Max dKS \cr
  \hline\hline 

	0 & 0.028&0.035\cr
	55 &0.036&0.043\cr
	110 &0.034&0.041\cr
	220 & 0.033&0.052\cr
	275 &0.029&0.068\cr
	440 &0.036&0.137\cr
	550 &0.030&0.173\cr
	825 &0.088&0.296\cr
	880 &0.107&0.351\cr
	1100 &0.220&0.560\cr

  \hline\hline
  \end{tabular}
  \end{center}
\end{minipage}
\end{table*}

\begin{table*}
\begin{minipage}{170mm}
  \begin{center}
  \caption[]{Probability of Extrema}
  \label{tab1}
  \begin{tabular}{c|c|c|c}\hline
  scale (') & Prob($T$$\ge$ $T_{max}COBE$) $\%$& Prob($T \le T_{min}COBE$)$\%$ & 
Prob($\vert (T/ \sigma_{pix})\vert \ge 
\vert (T/ \sigma_{pix})_{max}COBE \vert $)$\%$ \cr
 \hline\hline 

	108 &84.2&80.6& 96.8\cr
	324 &38.4&86.6 &82.6\cr
	432 &15.4&90.6 &41.6\cr
	648 &24.0&84.4&44.6\cr
	864 &67.6&62.4&78.2\cr

  \hline\hline
  \end{tabular}
  \end{center}
\end{minipage}
\end{table*}

\begin{table*}
\begin{minipage}{170mm}
  \begin{center}
  \caption[]{Probability of Areas outside the $99\%$ and the $95\%$}
  \label{tab1}
  \begin{tabular}{c|c|c|c|c}\hline
  scale (') & COBE pix out $99\%$ & Prob(pix out $99\% \ge$ COBE) $\%$ & 
COBE pix out $95\%$ & Prob(pix out $95\% \ge$ COBE) $\%$\cr
  \hline\hline 

	108 & 87&0.0&359&2.0\cr
	324 & 44&37.0&187&82.4\cr
	432 & 34&48.4&161&77.6\cr
	648 & 18&45.6&50&89.2\cr
	864 & 0&100.0&22&82.0\cr

  \hline\hline
  \end{tabular}
  \end{center}
\end{minipage}
\end{table*}


\section{Results and Conclusions}

We have applied the statistical tests described in the previous section to
the COBE-DMR data. 500 Gaussian simulations were performed to obtain
the confidence intervals for each test as well as the distribution functions
at each pixel. The results are summarised in figures 2 and 3 and tables 1, 2 and 3. For the tests 
in the first category, corresponding to figures 2 and 3, 
all but one of the COBE-DMR values are inside the $95\%$ confidence 
interval. In the case of $C_{2,1}$ 
between scales $55'$ and $110'$, the value for the COBE-DMR is inside the
$99\%$ confidence interval. Since the correlations are calculated between
scales that do not differ from each other by a constant amount, one can observe
an oscillating behaviour for $C_{2,1}$ and $C_{2,2}$. Some peaks occur
when the correlation is calculated between two scales differing
by a small amount. 
Also part of the first category tests, the
Kolmogorov-Smirnov distance obtained from the COBE-DMR data is, at all
scales, smaller than the maximum found for the 500 Gaussian simulations (Table 1). 
In the second category of tests (Tables 2 and 3), 
only the probability of finding a number
of pixels outside the $99\%, 95\%$ confidence intervals, greater or equal
the number of  COBE-DMR pixels found outside those intervals, at the 
smallest scale ($108'$), is very low. However, noise might be dominating at this scale. Therefore, no obvious deviation from Gaussianity can be noted from
any of the tests presented in this work.

The 
spherical Mexican Hat wavelet is introduced in this paper 
as a tool to study the information at different scales in the COBE-DMR data.
The only other wavelet on the sphere previously used to analyse 
CMB data is the spherical Haar wavelet (Tenorio et al. 1999, 
Barreiro et al. 2000). The main difference between these two wavelets is
the geometrical property of isotropy that characterises the Mexican Hat. 
The Haar wavelet however biases certain directions. As a consequence, 
orientation
of the data has to be taken into account in case this last wavelet is 
used in the analysis. The formalism of projecting the Mexican Hat 
(defined on the plane) on the surface of the sphere was presented in 
Section 2. Specific details related for example to the implementation on the HEALPix
pixelitation will be given in Mart\'\i nez-Gonz\'alez et al. (2001).

In summary, several statistical tests have been performed in wavelet space. They are 
divided in two categories. The first one includes tests performed globally
on the wavelet coefficient maps. Tests in the second category take into
account local properties at each scale. Temperature fluctuation distributions
are computed from Gaussian simulations at each pixel at each scale. The 
local dispersion is considered in the statistical tests applied. These
methods have not been applied in any of the previous works. 
From all 
of them we can conclude that the analysed COBE-DMR data are compatible with
Gaussianity. Skewness, kurtosis and $C_{2,2}$ were also tested by
Barreiro et al. 2000 with the spherical Haar wavelet. There is no 
indication from this work that one of the two wavelets is more optimal
for detecting non-Gaussian features in the COBE-DMR data. However, the
Mexican Hat wavelet is more localized in wavelet space than the
Haar wavelet. Moreover, a comparison of the performance on the
sphere of the
two wavelets shows, that the spherical Mexican Hat wavelet is 
more optimal than the spherical Haar wavelet for detecting certain 
non-Gaussian features (Mart\'\i nez-Gonz\'alez et al. 2001).

\section*{Acknowledgments}

The authors would like to thank Belen Barreiro and Luis Tenorio
for helpful comments.
LC, JLS EMG and FA thank Spanish DGESIC Project no.
PB98-0531-c02-01 for partial support. LC, JLS, EMG and JEG 
thank FEDER Project no. 1FD97-1769-c04-01, for financial support. 
We also thank the European Commission for partial finantial 
support through the RTN CMBNET contract HPRN-CT-2000-00124.

\end{document}